\documentclass[10pt,letterpaper]{article}

\usepackage[final]{cogsci}

\usepackage{booktabs}
\usepackage{dsfont}
\usepackage{comment}
\usepackage{graphicx}
\usepackage{subcaption}
\usepackage{xcolor}
\usepackage[
  style=apa,
  natbib=true,
  annotation=false,
]{biblatex}
\begin{document}

\newcommand{\R}{\mathcal{R}}
\newcommand{\A}{\mathcal{A}}
\newcommand{\C}{\mathcal{C}}
\newcommand{\fe}{$e_{\phi}$}
\newcommand{\ce}{$e_C$}
\newcommand{\TODO}[1]{{\textbf{TODO:}}}
\title{What Do You \emph{Think} I Think? Accounting for Human Beliefs\\Using Second-Order Theory of Mind\\
}
\author{\mbox{Patrick Callaghan (callaghan@cmu.edu)}}
\author{\mbox{Reid Simmons}}
\author{\mbox{Henny Admoni}}
\affil{Robotics Institute, Carnegie Mellon University}

\maketitle

\begin{abstract}

Discrepancies between an agent's actual knowledge and what a person thinks the agent knows can hinder interactions. If an agent could detect such discrepancies, it could provide feedback to account for them and improve current and future interactions. Using the I-POMDP as a framework for a second-order Theory of Mind (ToM-2), this work endows an agent with the ability to model the evolution of a person's erroneous beliefs about an agent and the cognitive biases and heuristics (CBH) from which they arise. In doing so, the agent can detect when CBH might be at play during an interaction and adaptively generate feedback that accounts for them. An in-person user study shows how a ToM-2 learner can account for the effects of a teacher’s CBH to significantly improve the informativeness of teacher actions, and subjective results suggest people find the ToM-2 learner's feedback more useful.

\end{abstract}

\section{Introduction}\label{sec:intro}

Intelligent agents can support people in a variety of interactive scenarios in the home or the workplace, but for the best chance at having successful interactions, people should have an accurate model of what the agent is capable of or likely to do. However, people employ \emph{cognitive biases and heuristics} (CBH) that may (incorrectly) influence their expectations for, and understandings of, the agent~\citep{tverskyJudgmentUncertaintyHeuristics1974,toddSimpleHeuristicsThat,gilovichHeuristicsBiasesPsychology2002}. In this work, we aim to improve human-agent interactions by enabling an agent to model the influence CBH can have on a person's beliefs about the agent so it can take actions to account for CBH's impact on said beliefs.

Whereas Theory of Mind (ToM) refers to one's ability to infer another's beliefs from their actions~\citep{premackDoesChimpanzeeHave}, \emph{second-order} Theory of Mind (ToM-2) is the ability to infer an agent's beliefs about another agent's beliefs---a skill people recruit to facilitate interactions with one another~\citep{astingtonTheoryMindEpistemological2002, paperaDevelopmentSecondorderTheory2019, clarkGroundingCommunication1991}.
Prior work in human-AI interaction investigates the utility of endowing agents with a ToM or ToM-2~\citep{abriniProceedings1stWorkshop2025,doshiModelingHumanRecursive2012,hanLearningOthersIntentional2018,yuanSituBidirectionalHumanrobot2022,zhangAutoToMAutomatedBayesian2025,brooksBuildingSecondOrderMental2019}, and while valuable contributions, we wish to learn an approximation of the human's model of the agent to generate agent actions that positively influence the human's true model and their downstream behaviors. Additionally, we specifically aim to do so by modeling and accounting for ways CBH can influence what people \emph{think} an agent thinks.

Our key insight is that a ToM-2 enables the agent to explicitly model a person's beliefs of the agent's beliefs during an interaction such that the agent can adapt its behavior to account for erroneous beliefs and their possible causes. To investigate this idea, we leverage the nested modeling afforded by an Interactive POMDP (I-POMDP) to maintain explicit models of a person and the effects that CBH could have on their perceptions of the agent~\citep{gmytrasiewiczFrameworkSequentialPlanning2005}. In doing so, the agent can approximate the impact that different CBH might have upon a person's inferences of agent beliefs. As a result, the agent can identify discrepancies between those approximations and its actual knowledge, as well as the candidate causes of those discrepancies. In turn, the agent can adaptively generate feedback which aims to aid the person's reasoning and inference processes. 

\begin{figure}
    \begin{center}
        \includegraphics[width=0.5\textwidth]{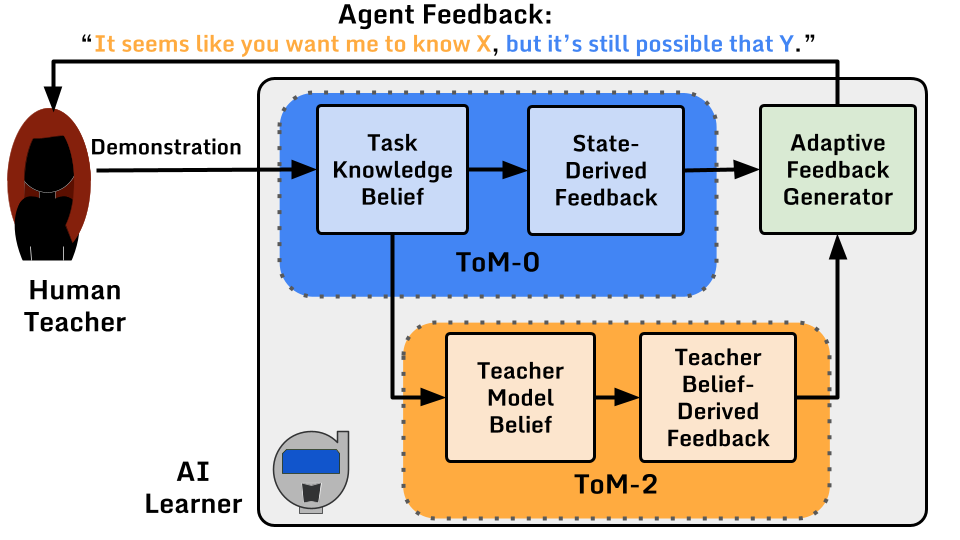}
        \end{center}
    \caption{An AI learner with a zeroth-order Theory of Mind (blue) maintains state-based beliefs informed by a human teacher's demonstrations but does not model human mental states. It provides feedback only with respect to state. ToM-2 augments the learner's belief with a model of what its teacher thinks the learner knows (orange) to provide additional feedback with respect to approximated teacher beliefs.}
    \label{fig:1}
\end{figure}

We contrive a teacher-learner domain to model how the \emph{Representativeness Heuristic} and \emph{Confirmation Bias} might affect a teacher's perceptions of an AI learner and their pedagogical reasoning. Furthermore, we design agent feedback that adaptively optimizes the information it communicates based on its beliefs and its approximation of the teacher's beliefs. With this formulation, we conduct an in-person user study and showcase:
\begin{itemize}
    \item A novel I-POMDP-based framework that enables an agent to approximate when and how CBH affects a human's behavior and beliefs;\label{takeaway:approx}
    \item This model enables an agent to generate feedback that accounts for a person's CBH upon being detected; and \label{takeaway:inference}
    \item Empirical evidence that this feedback results in improved interactions between humans and agents.\label{takeaway:confident}
\end{itemize}
 
\section{Preliminaries}\label{sec:preliminaries}
\begin{figure}[tb]
    \centering
    \includegraphics[scale=0.25]{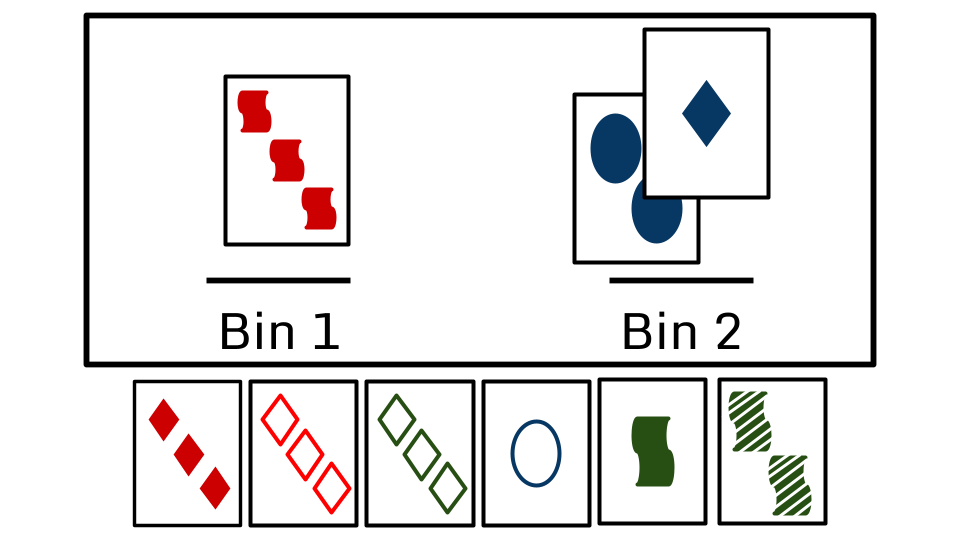}
    \caption{\textbf{Proof-of-Concept Domain}: Multi-featured cards are sorted into bins according to a rule.}
    \label{fig:domain}
\end{figure}

While our work is intended to generalize beyond the teacher-learner context, work in pedagogical reasoning is especially relevant given our domain of study~\citep{chenHierarchicalBayesianModel2024}. Shafto et al. formalize teaching as a recursive reasoning process and assume a teacher selects its data with helpful intent~\citep{shaftoRationalAccountPedagogical2014}. Our work compliments this formal representation using an alternative method for recursive reasoning from the learner's perspective while aiming to investigate how models of human CBH might affect their ability to select helpful data. Additionally, Rafferty et al. showcase how POMDPs enable the planning of teaching actions when faced with different cognitive models~\citep{raffertyFasterTeachingPOMDP2016}. While important work and similar to ours in multiple ways, our focus is on how a ToM-2 enables an AI learner to model and account for a teacher's potential CBH when selecting learner actions.

\textbf{Domain} We designed a turn-based teaching task to study how a ToM-2 can enable an agent to model, detect, and account for a person's CBH during an interaction (Figure~\ref{fig:domain}). In this task, a ``rule'' governs how multi-featured cards are sorted into two bins. There are four Classes of card features (Color, Fill, Number, Shape), each with three possible Feature Values (e.g., Blue, Green, Red). A rule consists of one Class from which two Feature Values are assigned to bins one and two respectively. Crucially, the third Feature Value can go in \emph{both} bins, thereby introducing ambiguity which the teacher must also resolve. For example, one rule could be: (Class: Color. Bin 1 Value: Red. Bin 2 Value: Blue). Here, Green cards can be placed in either bin. Going forward, this rule will serve as a running example to provide intuition where helpful.

In a single teaching round, the teacher sorts a card into a bin and the agent responds with a statement pertaining to its current belief over rule features (detailed in the Approach section). This process continues until the teacher determines that the agent knows the true rule with absolute certainty and attempts to end the session. If the learner knows the rule, the session ends; otherwise, the session recommences until the teacher's next attempt to end it.

\textbf{Model} An Interactive Partially Observable Markov Decision Process (I-POMDP) for an agent with a level-$L$ Theory of Mind (ToM-$L$) comprises the tuple: $\langle IS_L, A, T, O, R, \Omega \rangle$ with $[A]$ction and $[O]$bservation spaces, $[T]$ransition and $[R]$eward functions, and observation function $\Omega$~\citep{gmytrasiewiczFrameworkSequentialPlanning2005}. Unlike a traditional POMDP, the I-POMDP models an \emph{interactive state} space $IS_L$ in which each $is_L \in IS_L$ comprises a state $s$ and a model of another agent $m_{L-1}$, which itself is a level-$(L-1)$ I-POMDP.

In the card-sorting domain, each state is a sorting rule. As such, each of a ToM-2 $[\A]$gent's $is_2 \in IS_2$ comprises a rule and a model $m_{H, 1}$ of the $[H]$uman as a ToM-1 I-POMDP. 

A POMDP's belief space is already intractable; the I-POMDP's nested belief space only exacerbates the problem. Whereas prior work accounts for intractability through efficient solving techniques~\citep{hanIPOMDPNetDeepNeural2019,doshiMonteCarloSampling2009,schwartzOnlinePlanningInteractivePOMDPs2022}, we use the I-POMDP strictly as a framework to track second-order beliefs. As such, \textbf{we do not solve the I-POMDP} and thereby avoid its costliest aspect. Beliefs, however, still need to be updated, and ToM-2 belief updates modify the standard POMDP as: 
\begin{equation}\label{eq:tom2beliefhumanaction}
    \begin{split}
        b_{\A}(is^t) = \Omega_{\A}(o_{\A}^t |is^t) \sum\limits_{is^{t-1}} b_{\A}^{t-1} T(s^t, s^{t-1}) P(b_{H}^t|b_{H}^{t-1}, a_{H}^{t-1}),
    \end{split}
\end{equation}
where the final term encodes recursive updates to the agent's ToM-1 and ToM-0 beliefs\footnote{Equation~\eqref{eq:tom2beliefhumanaction} abbreviates a more substantial representation whose derivation can be found in the original I-POMDP paper~\citep{gmytrasiewiczFrameworkSequentialPlanning2005}.}.

\section{Approach}\label{sec:approach}

Prior work with I-POMDPs largely focuses on predicting and responding to actions another agent will take in a multi-agent, non-collaborative setting~\citep{hanIPOMDPNetDeepNeural2019,rathnasabapathyExactSolutionsInteractive2006,doshiModelingHumanRecursive2012}. In contrast, we use I-POMDPs to represent a human's beliefs about an agent's beliefs so an agent can generate feedback meant to influence those beliefs and the human's subsequent actions.

To achieve this goal, we fashion the agent's ToM-2 architecture with two principal functions: \emph{CBH detection} and \emph{adaptive feedback generation}.

\subsection{Detecting Cognitive Biases and Heuristics}\label{sec:detect}
 
The I-POMDP framework enables the agent to maintain beliefs over multiple human models $m_{H, 1} \in M_{H, 1}$ that encode $[\C]$BH mechanisms as functions:
\begin{equation}
    \C(i) \quad i \in \{ IS, A, T, R, O, \Omega \},
\end{equation}
where $\C$ applies some systematic modification to the underlying representation of its input. As a starting example, we approximate how a human's CBH affects their observations (i.e., $P(o|is, \C)$), and in this work, we focus on the \emph{Representativeness Heuristic} (RH) and \emph{Confirmation Bias} (CB).

People use the RH to assess an event's probability by estimating its similarity to their subjective prior models, and the Local Thinking (LT) formalization models the degree to which a person relies upon the RH as the limited number of features they attend to when performing this assessment~\citep{ErrorsProbabilisticReasoning2019, gennaioliWhatComesMind2010}. We operationalize LT by augmenting $\Omega$ with a weighted cosine similarity:
\begin{equation}\label{eq:representativeness}
    \begin{split}
        \Omega_\C(is, o) &=  P\big(o |is, \cos\big(\tilde{\mathbf{\Phi}}(is),\tilde{\mathbf{\Phi}}(o)\big); \beta \big),
    \end{split}
\end{equation}
\noindent
where $\beta$ modulates the cosine similarity's effect on the observation likelihood\footnote{$\beta$ also distinguishes the human models maintained by the ToM-2 agent.}. In the card-sorting domain, LT will cause teachers to focus on the Feature Values that are semantically similar to a rule while neglecting the semantically dissimilar counterfactual examples needed to logically eliminate other candidate rules. In our running example, this means people would continue to sort Blue and Red cards while overlooking the other Class' Feature Values they've yet to eliminate.

CB causes people to seek events that align with their prior beliefs, preferences, or expectations~\citep{nickersonConfirmationBiasUbiquitous1998}. In our framework, CB and the RH are functionally identical but operationally distinct: Whereas the RH affects a human's reasoning over what an agent infers from human actions (i.e., $\mathbb{E}_H(\Omega_\A(a_H))$), CB affects a human's inferences from observing the agent's actions (i.e., $\Omega_H(a_\A)$). In our domain, CB would cause teachers to expect agent feedback that explicitly addresses the rule they teach; in our running example, this means teachers begin to expect learner statements about Red distinguishing Bin 1 and Blue distinguishing Bin 2 as opposed to, say, Diamond's relationship to either bin.

These modifications alter the nested belief updates in Equation~\ref{eq:tom2beliefhumanaction}. In turn, a ToM-2 agent can detect a person's tendency to rely on the RH and CB by assigning more probability mass to those human models whose expected actions align with those of the actual human:
\begin{equation}
    \mathds{1}(\mathbb{E}_\A[a_H|m_{H, 1}] \equiv o_\A).
\end{equation}
\subsection{Adaptive Feedback Generation}\label{sec:feedback_generation} 
As in prior research, we assume people expect two general flavors of agent feedback: That which seeks information, and that which reveals it~\citep{habibianHereWhatVe2021}. While the agent could present feedback for both feedback types, this strategy risks excess cognitive burden for the human~\citep{cheyettePeopleSeekEasily, amershiPowerPeopleRole2014}. As such, we optimize for information gain by default and additionally for transparency when a person's behavior indicates they have erroneous beliefs about the agent's beliefs that we assume are manifested by CBH. Each optimization generates a different type of statement.

\subsubsection{\textbf{\textcolor{blue}{Confidence Statements (CS)}}}\label{sec:CS}
A CS conveys the agent's uncertainty over a rule feature which, if resolved, would yield large information gain for the learner. It consists of two clauses: the \emph{confidence expression} (\ce) and the \emph{feature expression} (\fe). First, the agent generates a set of viable \fe \ for each rule $s$ by finding the \fe \ consistent with that rule's features $\phi_s \in \mathbf{\Phi}(s)$. Two viable \fe \ for our running example include ``Bin 1 Value is Red,'' and ``Bin 1 Value is Green.''
Then, the agent selects an \ce \ from the set: ``I know,'' ``I think,'' or ``I'm unsure if,'' meant to encode a descending expression of confidence in the selected \fe.
The closer to uniform the belief over the features $\mathbf{\Phi}(s)$, the likelier the agent is to select the ``I'm unsure if'' statement; the closer to a singleton belief, the likelier it is to select ``I know.'' 

\subsubsection{\textbf{\textcolor{orange}{Understanding Statements (US)}}}\label{sec:US}
A US is meant to ``nudge'' teachers out of CB-induced teaching actions by revealing what the agent knows about a Feature Value the person likely intends to teach. We generate a US when human behavior indicates a discrepancy between the agent's belief of a rule's Feature Value $\phi_s$ and the beliefs the human thinks it has:
\begin{equation}\label{eq:cogbias_detection}
    \begin{split}
        \mathcal{E}_{\A, 0}(\phi) &= ||d(b_{\A, 2}(\phi_s), \ b_{\A, 0}(\phi_s))|| \ \ \ \forall b_{\A, 0} \in M_{H, 1}, \\
    \end{split}
\end{equation}
where $d(\cdot, \cdot)$ is a linear distance computation. We then use this $[\mathcal{E}]$rror term to compute optimal time $t$ actions:
\begin{equation}\label{eq:cogbias_reward}
    \begin{split}
        R_{\A}(US; \ \phi_s) &= \\ 
        b_{\A, 2}(\phi_s) &\sum\limits_{m_{H, 1} \atop \in IS_{\A, 2}} b_{\A, 2}(m_{H, 1}) \sum\limits_{b_{\A, 0} \atop \in m_{H, 1}} b_{\A, 0}(\phi_s) \mathcal{E}_{\A, 0}. \\
    \end{split}
\end{equation}
When the reward value exceeds an empirically-chosen cost threshold, the agent addresses this discrepancy in its $US(\phi_s)$.

If the agent doesn't detect erroneous beliefs, it presents a single CS assuming the teacher can use the information without CBH to select an optimal teaching action. If it does detect erroneous beliefs, the agent presents a (US+CS) tandem.

In the card-sorting domain, the ToM-2 learner's action-space comprises statements about Classes and Feature Values. For example, a CS about the rule's Class might be, ``I think the Class is Shape,'' and feedback comprised of (US+CS) statements would take the form, ``\textcolor{orange}{It seems like you want me to know the Class is Shape}, but \textcolor{blue}{it's still possible that the Class is Fill.}'' Statements are limited to one rule component so as to present the teacher with a reasonable inference challenge.

The teacher's action-space comprises card placements consistent with the rule they teach and a ``Terminate'' option used when they believe the learner knows the rule.

\section{Evaluation}\label{sec:evaluation}
\begin{table*}[]
\resizebox{\textwidth}{!}{%
\begin{tabular}{@{}lllllll@{}}
\toprule
\begin{tabular}[c]{@{}l@{}}Learner + \\ Teacher \end{tabular} & 
  \multicolumn{1}{c}{$t = 1 $} &
  \multicolumn{1}{c}{$t = 5 $} &
  \multicolumn{1}{c}{$t = 7 $} &
  \multicolumn{1}{c}{$t = 23 $} &
  \multicolumn{1}{c}{$t = 31 $} &
  \begin{tabular}[c]{@{}l@{}}Final Belief of Ground \\Truth Teacher Model\end{tabular} \\ \midrule
\textbf{\begin{tabular}[c]{@{}l@{}}ToM-0 +\\  No CBH\end{tabular}} &
  \begin{tabular}[c]{@{}l@{}}“I’m unsure if the\\ Class is Color.”\end{tabular} &
  \begin{tabular}[c]{@{}l@{}}“I’m unsure if Bin 1 is\\Squiggle.”\end{tabular} &
  \begin{tabular}[c]{@{}l@{}}“I know Bin 2 is Oval.”\end{tabular} &
  \textbf{N/A} &
  \textbf{N/A} &
  \textbf{N/A} \\ \cmidrule(l){2-7} 
\textbf{\begin{tabular}[c]{@{}l@{}}ToM-2 +\\ No CBH\end{tabular}} &
  \begin{tabular}[c]{@{}l@{}}“I’m unsure if the\\ Class is Color.”\end{tabular} &
  \begin{tabular}[c]{@{}l@{}}“It seems like you want me\\to know Bin 1 is Squiggle,\\but it still could be Diamond.”\end{tabular} &
  \begin{tabular}[c]{@{}l@{}}“I know Bin 2 is Oval.”\end{tabular} &
  \textbf{N/A} &
  \textbf{N/A} &
  \textbf{0.9} \\ \midrule
\textbf{\begin{tabular}[c]{@{}l@{}}ToM-0 +\\  CBH\end{tabular}} &
  \begin{tabular}[c]{@{}l@{}}“I’m unsure if the\\ Class is Color.”\end{tabular} &
  \begin{tabular}[c]{@{}l@{}}“I think the Class is Shape.”\end{tabular} &
  \begin{tabular}[c]{@{}l@{}}“I think the Class is Shape.”\end{tabular} &
  \begin{tabular}[c]{@{}l@{}}“I think Bin 1 is Diamond.”\end{tabular} &
  \begin{tabular}[c]{@{}l@{}}“I think Bin 2 is Oval.”\end{tabular} &
  \textbf{N/A} \\ \cmidrule(l){2-7} 
\textbf{\begin{tabular}[c]{@{}l@{}}ToM-2 +\\  CBH\end{tabular}} &
  \begin{tabular}[c]{@{}l@{}}“I’m unsure if the\\ Class is Color.”\end{tabular} &
  \begin{tabular}[c]{@{}l@{}}“It seems like you want me\\ to know the Class is Shape,\\but it still could be Color.”\end{tabular} &
  \begin{tabular}[c]{@{}l@{}}“It seems like you want me\\to know the Class is Shape,\\but it still could be Fill.”\end{tabular} &
  \begin{tabular}[c]{@{}l@{}}“It seems like you want me\\to know Bin 1 is Diamond,\\ but it still could be Squiggle.”\end{tabular} &
  \begin{tabular}[c]{@{}l@{}}“It seems like you want me\\ to know Bin 2 is Oval,\\but it still could be Squiggle.”\end{tabular} &
  \textbf{0.99}
  \\ \bottomrule
\end{tabular}%
}
\caption{\textbf{Qualitative Comparison of ToM-0 and ToM-2 Feedback with Simulated Teachers:} When the teacher has no CBH, learner feedback is comparable. With CBH, the ToM-2 learner generates US and CS, but the ToM-0 learner can only generate CS. Moreover, the ToM-2 learner identifies the correct rule and teacher model, which a ToM-0 learner cannot do (last column). }
\label{tab:sim}
\end{table*}

To evaluate our ToM-2 framework, we conducted simulated experiments for qualitative validation in addition to a formal user study. Note that in the teacher-learner dyad, a first-order learner is indistinguishable from a zeroth-order learner: Zeroth-order models encode the rules being taught; first-order models encode the teacher's belief of the rule being taught (which is always known to the teacher and thus is redundant). As such, a ToM-1 baseline is omitted in our evaluations.

\subsection{Qualitative Validation}

We simulated teachers to examine differences in how ToM-0 and ToM-2 learners adaptively generate feedback (Table~\ref{tab:sim}). In these experiments, the simulated human taught the rule: (Class: Shape. Bin 1 Value: Diamond. Bin 2 Value: Oval). As expected, the interactions are hardly distinguishable with an optimal teacher. However, when CBH is injected into the teacher model, teaching sessions become longer due to the teacher's reliance upon the RH. The ToM-2 learner adapts to the model's teaching by recruiting its (US+CS) feedback to address their likeliest intentions given the teacher's actions.

\subsection{User Study}\label{sec:study}
We conducted an in-person, within-subjects user study in which participants used a GUI to teach a virtual agent. The study was approved by CMU's Internal Review Board.

\subsubsection{Participant Demographics} We recruited 35 individuals aged 18 or older from the institution's pool of volunteer participants and performed analysis using data from 30 (three removed for failure to understand the teaching task; two for a GUI malfunction). Of the 30 participants, 19 identified as female, 10 as male, and one as non-binary/non-conforming. Mean age was 26.25 (S.D. 8.1). 20 participants had at least a Bachelor's degree; 8 had some college but no degree.

\subsubsection{Experimental Conditions} Participants experienced three conditions primarily distinguished by the learner's ToM order (zeroth or second) and secondarily distinguished by the learner's feedback policy. Since US are generated via discrepancies between the learner's actual belief and its approximation of what the teacher thinks it believes, a ToM-0 agent is incapable of generating US. This shortcoming risks unfair comparison since more information is present in the ToM-2 feedback, so the ToM-0 baselines were granted the ability to provide two CS either when the ToM-2 model would have provided (US+CS) feedback (the ``ToM-0'' condition) or at random time steps (the ``ToM-0 Random'' condition).

\subsubsection{Hypotheses}
\begin{itemize}
    \item[] \textbf{H1:} The ToM-2 learner will elicit more efficient teaching.
    \item[] \textbf{H2:} Two-statement feedback induced by the ToM-2 model will elicit more informative teaching.
    \item[] \textbf{H3:} People will feel more confident with the ToM-2 learner.
\end{itemize}
\subsubsection{Measurements} We tracked multiple metrics related to teaching efficiency and subjective experience:
\begin{itemize}
    \item[] \textbf{M1:} Number of cards teacher places in each session.
    \item[] \textbf{M2:} Number of cards teacher places between the time step the agent learned the rule and when they end the session.
    \item[] \textbf{M3:} Number of times teacher attempts to end the session.
    \item[] \textbf{M4:} Time teacher takes deliberating next card placement.
    \item[] \textbf{M5:} Time teacher takes to complete a teaching session.
    \item[] \textbf{M6:} Relative information gain of teacher card placements.
    \item[] \textbf{M7:} After each piece of agent feedback, participants answer a Likert prompt about their confidence that the agent is learning from their teaching (Confidence from Feedback). If a participant attempts to end the teaching session, they answer a Likert prompt about their confidence that the agent knows the rule (Termination Confidence). After each teaching session, participants rate the relevance of the agent's feedback and their confidence that they understood the agent's rule knowledge throughout the session (Confidence in Agent's Understanding).
    \item[] \textbf{M8:} After each session, participants rate the pleasantness of their teaching experience on a Likert scale. At the study's conclusion, they identify which agent was easiest to teach.
\end{itemize}
\subsubsection{Procedure} Participants were fully counterbalanced between all orderings of learner conditions. After receiving spoken instructions and signing a consent form, participants engaged with an on-screen tutorial detailing the sorting task and the GUI. They then briefly practiced interacting with the GUI by placing cards in bins and receiving feedback from the virtual ToM-0 agent. Then the three teaching sessions commenced. Prior to concluding the study, participants completed a demographics questionnaire.

A monetary bonus incentivized participants to teach as efficiently as possible. If they ended the teaching session at the time step when the agent learned the rule, they earned an extra dollar. The bonus decreased proportional to the number of time steps between when the agent learned the rule and when the teacher attempted to end the teaching session.

\subsubsection{Results}
We conducted one-way repeated measures ANOVAs for each of the M1-M6 metrics; if the sphericity assumption was violated, we applied a Greenhouse-Geisser correction. Post-hoc, we applied pairwise two-sided t-tests to test for differences between conditions, though we used Wilcoxon sign ranked tests if t-test assumptions were violated. 

\begin{figure*}
    \captionsetup[subfigure]{labelformat=empty}
    \centering
    \begin{subfigure}[t]{0.95\columnwidth}
        \centering
        \includegraphics[height=2.2in]{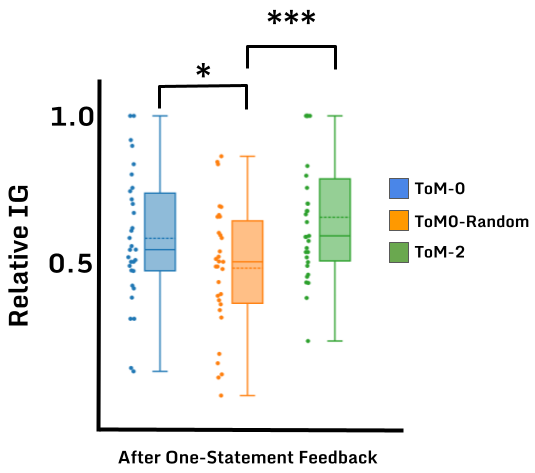}
        \caption{}
        \label{fig:rel_ig_0_box}
    \end{subfigure}%
    \hfil
    \begin{subfigure}[t]{0.95\columnwidth}
        \centering
        \includegraphics[height=2.2in]{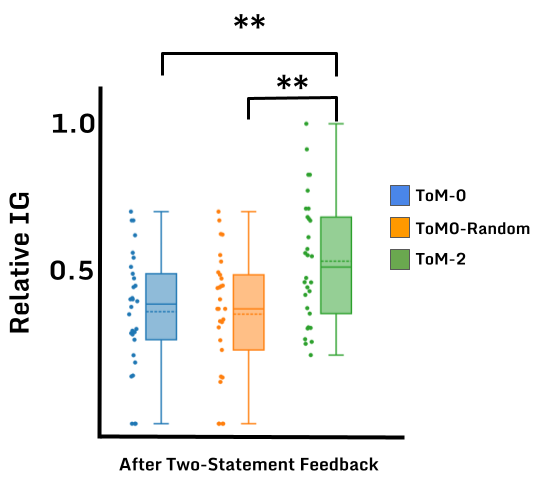}
        \caption{}
        \label{fig:rel_ig_chopped_box}
    \end{subfigure}
    \caption{\textbf{Mean Relative Information Gain: } The IG of the card placed by the human normalized by the IG of that time step's optimal card. \textbf{Left: }When holding feedback content constant, the ToM-0 and ToM-2 learners elicit more informative teaching due to the ToM-2 learner's adaptive feedback process. \textbf{Right: }The content of the ToM-2 learner's two-statement feedback elicits more informative teaching.}
\label{fig:rel_ig_box}
\end{figure*}
\textit{Relative Information Gain (IG)} To investigate the usefulness of the ToM-2 agent's CBH detection and adaptive feedback generation, we compared the informativeness of cards placed after one- and two-statement feedback across the conditions. To disambiguate the effects of content (i.e., \emph{what} the agent says) and timing (i.e., \emph{when} the agent adapts), we performed two evaluations. The first investigates timing: The ToM-2 learner's ability to detect erroneous beliefs and then adapt the form of its feedback. This effect can be examined after one-statement feedback since this feedback is always a CS, yet the ToM-2 learner's mechanism for choosing to provide one- or two-statement feedback is still operational. As seen in Figure~\ref{fig:rel_ig_0_box}, we found statistically significant differences between the conditions ($F(2, 58) = 7.163, \ p < 0.002$), and post-hoc analysis found statistically significant differences between ToM-2 and ToM-0 Random conditions ($t(29) = 3.851, \ p < 0.0006$) as well as between the ToM-0 and ToM-0 Random conditions ($t(29) = 2.358, \ p < 0.025$).

The second evaluation examines content (the utility of the ToM-2 learner's US), however, a pre-processing step was necessary. Since it takes multiple teaching actions for the ToM-2 learner to detect CBH, it always generates two-statement feedback later in the interaction when there are fewer cards that would yield large IG. In contrast, the ToM-0 Random learner can generate two-statement feedback early in the interaction when there are many cards that yield large IG. Thus, the IG after its two-statement feedback is positively biased.

To account for this imbalance, we identified the ceiling of the mean time step at which the ToM-0 and ToM-2 learners first presented two-statement feedback (which was 3) and used it as the starting time step for the ToM-0 Random condition's data. The result (Figure~\ref{fig:rel_ig_chopped_box}) reveals statistically significant differences after two-statement feedback $(F(2, 58) = 7.888, \ p < 0.0009)$, and post-hoc analysis finds statistical significance between the ToM-2 condition and each of the other conditions: $t(29) = 3.315, \ p < 0.0025$ (ToM-0), and $t(29) = 3.199, \ p < 0.003$ (ToM-0 Random).

\begin{figure}
    \centering
    \includegraphics[height=1.8in, width=0.99\columnwidth]{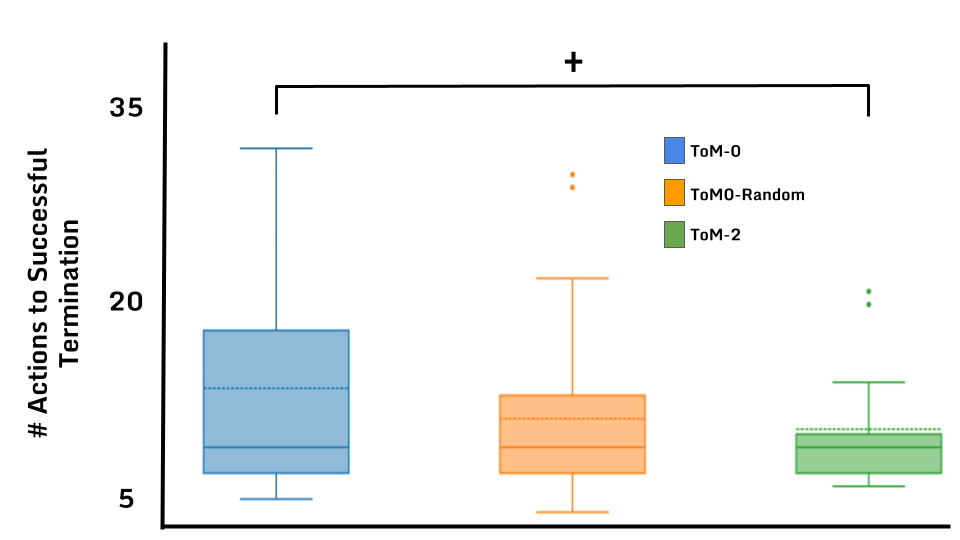}
    \caption{\textbf{Mean Teacher Action Count: } ToM-2 learners tend to elicit fewer teacher actions before teachers successfully end the session. Marginal significance indicated by +.}
    \label{fig:teacher_counts}
\end{figure}
\textit{Teaching Efficiency}
The ToM-2 model reduces the number of early attempts to end teaching sessions to a marginally significant degree $(F(2, 58) = 2.394, \ p < 0.100)$, but post-hoc analysis only found a marginally significant difference between the ToM-2 and ToM-0 conditions ($t(29) = 1.919, \ p < 0.065$). Similarly, we found a marginally significant difference in the number of teacher actions before successfully ending the teaching sessions ($F(2, 58) = 2.614, \ p < 0.082$), and post-hoc analysis only found a statistically significant difference between the ToM-2 and ToM-0 conditions ($t(29) = 2.108, \ p < 0.044$) (Figure~\ref{fig:teacher_counts}). There was no significance for the remaining metrics in any direction.

\begin{table}[]
\resizebox{\columnwidth}{!}{%
\begin{tabular}{@{}llll@{}}
\toprule
 & \multicolumn{1}{c}{ToM-0} & \multicolumn{1}{c}{\begin{tabular}[c]{@{}c@{}}ToM-0 Random\end{tabular}} & \multicolumn{1}{c}{ToM-2} \\ \midrule
\begin{tabular}[c]{@{}l@{}}Confidence in Agent's \\Understanding\end{tabular} & $2.45 \pm 1.02$ & $2.66 \pm 1.17$ &$2.90 \pm 0.939$   \\ \midrule
\begin{tabular}[c]{@{}l@{}}Confidence \\ from Feedback\end{tabular}               & $2.47 \pm 0.652$ & $2.15 \pm 0.529$ &$2.45 \pm 0.705$   \\ \midrule
\begin{tabular}[c]{@{}l@{}}Termination\\ Confidence\end{tabular}            & $3.18 \pm 0.774$ & $3.02 \pm 0.849$ &$3.23 \pm 0.721$   \\ \midrule
\begin{tabular}[c]{@{}l@{}}Teaching\\ Confidence\end{tabular}               & $2.69 \pm 0.850$ & $2.90 \pm 1.01$ &$2.83 \pm 0.759$   \\ \midrule
\begin{tabular}[c]{@{}l@{}}Feedback\\ Utility\end{tabular}                  & $3.14 \pm 0.693 (*)$ & $3.14 \pm 0.875 (*)$ &$\mathbf{3.62 \pm 0.622}$   \\ \midrule
\begin{tabular}[c]{@{}l@{}}Feedback\\ Relevance\end{tabular}                  & $3.03 \pm 1.02 $ & $2.90 \pm 0.817 $ &$3.17 \pm 0.658$   \\ 
\end{tabular}%
}
\caption{\textbf{Subjective Metrics:} Means and standard deviations of self-reported scores from the 30 study participants. Bold = statistically significant compared to row's starred entries.}

\label{tab:subjective}
\end{table}
\textit{Subjective Experience}
Table~\ref{tab:subjective} reports means and standard deviations of our subjective metrics. We only found statistical significance with respect to Feedback Utility ($\chi^2(1.93, 56.07) = 8.190, \ p < 0.0009$), where post-hoc analysis found a statistically significant difference between ToM-2 and ToM-0 ($t(29) = 3.067, \ p < 0.005$) and ToM-2 and ToM-0 Random ($t(29) = 3.042, \ p < 0.005$). Results were computed using a Friedman test and post-hoc t-tests.

\section{Discussion}\label{sec:discussion}
\paragraph{Objective Findings} The ToM-2 learner elicits more informative teaching from people (\textbf{H1} and \textbf{H2}). As compared to the ToM-0 and  ToM-0 Random baselines, the ToM-2 learner incurs greater relative IG after one- and two-statement feedback alike, but detailed analysis revealed the importance of the ToM-2 learner's capacity to detect and adapt to CBH, as well as to provide CB-targeted feedback via US. More specifically, when isolating the effects of feedback timing from feedback content, statistically significant results indicate that both features of the ToM-2 architecture help elicit informative teaching from humans. This result underscores the utility of the ToM-2 learner's US---and its detection of the human's erroneous beliefs---to teacher and learner alike. In terms of teaching efficiency (\textbf{H1}), the ToM-2 learner tends to cause people to successfully end teaching sessions sooner (Figure~\ref{fig:teacher_counts}).

We didn't find statistical significance or trends when comparing the number of excess cards teachers presented the agent (M2), deliberation time (M4), or total teaching session duration (M5), suggesting participants often identified a sufficient teaching strategy regardless of the learner they taught. Given the domain's simplicity, this result is hardly surprising. Indeed, many participants indicated they found it easier to teach certain agents based not on the learner but rather their understanding of how to teach. Moreover, while the relative IG results indicate the importance of feedback content, the similarity of each learner's feedback dampened noticeable differences and effects of the conditions. Study participants' written responses revealed as much, with many stating they found the learners equally easy to teach, indistinguishable in their interactions, or both. As such, the study reveals a need to increase the task and domain complexities to further analyze the benefits afforded by a ToM-2 model.

\paragraph{Subjective Findings} While teachers felt the ToM-2 learner's feedback was most useful ($p<0.001$), we failed to find support for \textbf{H3} in terms of teacher confidence. The same was true for relevance, ease of teaching the rule, and pleasantness/frustration (M7, M8). 

Collectively, the objective and subjective findings indicate an interesting result: People tend to teach a ToM-2 learner better, but they aren't necessarily aware it's happening. Perhaps there is an awareness that the ToM-2 learner provides feedback that helps the teachers resolve ambiguity and achieve the desired outcome more efficiently, but the results seem to indicate that this awareness is subconscious at most. As with the objective findings, a more complex task is a good next step to further study a ToM-2 agent's subjective benefits for people.

\section{Conclusion}\label{sec:conclusion}

We introduced a method that endows an agent with a second-order Theory of Mind. With it, the agent can detect approximately when and how cognitive biases and heuristics might cause a person to develop erroneous beliefs about the agent's beliefs and therby hinder the human's reasoning and inference processes. In turn, the agent can adaptively generate feedback which accounts for the person's CBH so that they might reason more efficaciously. An in-person user study of a teacher-learner scenario showcased how the timing and content of the agent's feedback afforded by its ToM-2 model elicits more informative teaching from humans, and subjective results indicate people found the ToM-2 learner's feedback to be more helpful to them as compared to ToM-0 baselines.

This work is not without limitation. The proof-of-concept domain does not encapsulate the complexity of real-world human-agent interactions; in the same vein, people are rife with biases that may influence human-agent interactions at any given time, yet we explore just two. Finally, the I-POMDP's computational complexity must be addressed if this approach is to scale to more complex task scenarios and human models.

\section{Acknowledgments}\label{sec:acknowledgments}

We would like to thank CMU's Center for Behavioral and Decision Research for enabling us to recruit participants from the Pittsburgh community, as well as the participants themselves. We also would like to thank the many lab mates for their feedback and support. This work was made possible in part due to the generosity of SoftBank Group Corporation and the Microsoft Corporation.

\printbibliography

\end{document}